\documentclass[twocolumn]{el-author}

\newcommand{\ua}{\uparrow}
\newcommand{\nc}{\newcommand}
\nc{\da}{\downarrow} \nc{\hc}{\hat{c}} \nc{\hS}{\hat{S}}
\nc{\bra}{\langle} \nc{\ket}{\rangle} \nc{\eq}{equation (\ref}
\nc{\h}{\hat} \nc{\hT}{\h{T}}\nc{\be}{\begin{eqnarray}}
\nc{\ee}{\end{eqnarray}}\nc{\rd}{\textrm{d}}\nc{\e}{eqnarray}\nc{\hR}{\hat{R}}\nc{\Tr}{\mathrm{Tr}}
\nc{\tS}{\tilde{S}}\nc{\tr}{\mathrm{tr}}\nc{\8}{\infty}\nc{\lgs}{\bra\ua,\phi|}\nc{\rgs}{|\ua,\phi\ket}
\nc{\hU}{\hat{U}}\nc{\lfs}{\bra\phi|}\nc{\rfs}{|\phi\ket}\nc{\hZ}{\hat{Z}}\nc{\hd}{\hat{d}}\nc{\mD}{\mathcal{D}}
\nc{\bd}{\bar{d}}\nc{\bc}{\bar{c}}\nc{\mc}{\mathcal}\nc{\ea}{eqnarray}\nc{\mG}{\mathcal{G}}\nc{\bce}{\begin{center}}
\nc{\ece}{\end{center}}
\date{1st December 2014}

\begin{document}

\title{Interference Cancellation trough Interference Alignment for Downlink of Cognitive Cellular Networks}

\author{May Moussa, Fotis Foukalas and Tamer Khattab}

\abstract{In this letter, we propose the interference cancellation through interference alignment at the downlink of cognitive cellular networks. Interference alignment helps the spatial resources to be shared among primary and secondary cells and thus, it can provide higher degrees of freedom through interference cancellation. We derive and depict the achievable degrees of freedom. We also analyse and calculate the achievable sum rates applying water-filling optimal power allocation.}

\maketitle

\section{Introduction}
Cognitive cellular networks concept will rely on the efficient interference management and one key-technology is the Interference Alignment (IA) \cite{Liu} \cite{Gollakota}. IA was initially proposed as   technique for increasing the degrees of freedom (DoF) by achieving interference-free signalling dimensions in multiple-input-multiple-output (MIMO) systems \cite{Jafar}. IA as cognitive radio application should allow secondary users (SUs) to exploit the dimensions that are unused by the primary users (PUs) \cite{Spec_Pool}. In this work, we provide the cancellation of two types of interferences in the downlink of a cognitive cellular network applying an IA technique. To our knowledge, downlink IA has been recently investigated in \cite{Downlink_IA} but not in a cognitive cellular network for interference cancellation. 

Both \textit{intra-cell} interference, which is caused by transmissions intended to the other users operating in the same cell, and \textit{inter-cell} interference, which is caused  by transmissions occurring in the other cells \cite{IC} are assumed. To be specific, based on the model proposed in ~\cite{Joint_IC}, we consider that the base station (BS) of the primary cell (pCell) has the knowledge of the channel state information (CSI) within its cell received from its own PUs, while the BS of the secondary cell (sCell) has the knowledge of the CSI of all the channels including the PUs. We provide the bounds of the considered IA technique providing the design of the signalling scheme. We depict the achievable DoF regions and explain how the interference cancellation is achieved through the proposed IA. Moreover, we provide the rate maximization for both PUs and SUs using water-filling optimal power allocation over the singular values of the produced channel matrices.

\section{Interference Cancellation through Interference Alignment}
\begin{figure}  
  \centering
  \includegraphics[width=0.4 \textwidth, height=0.24 \textheight]{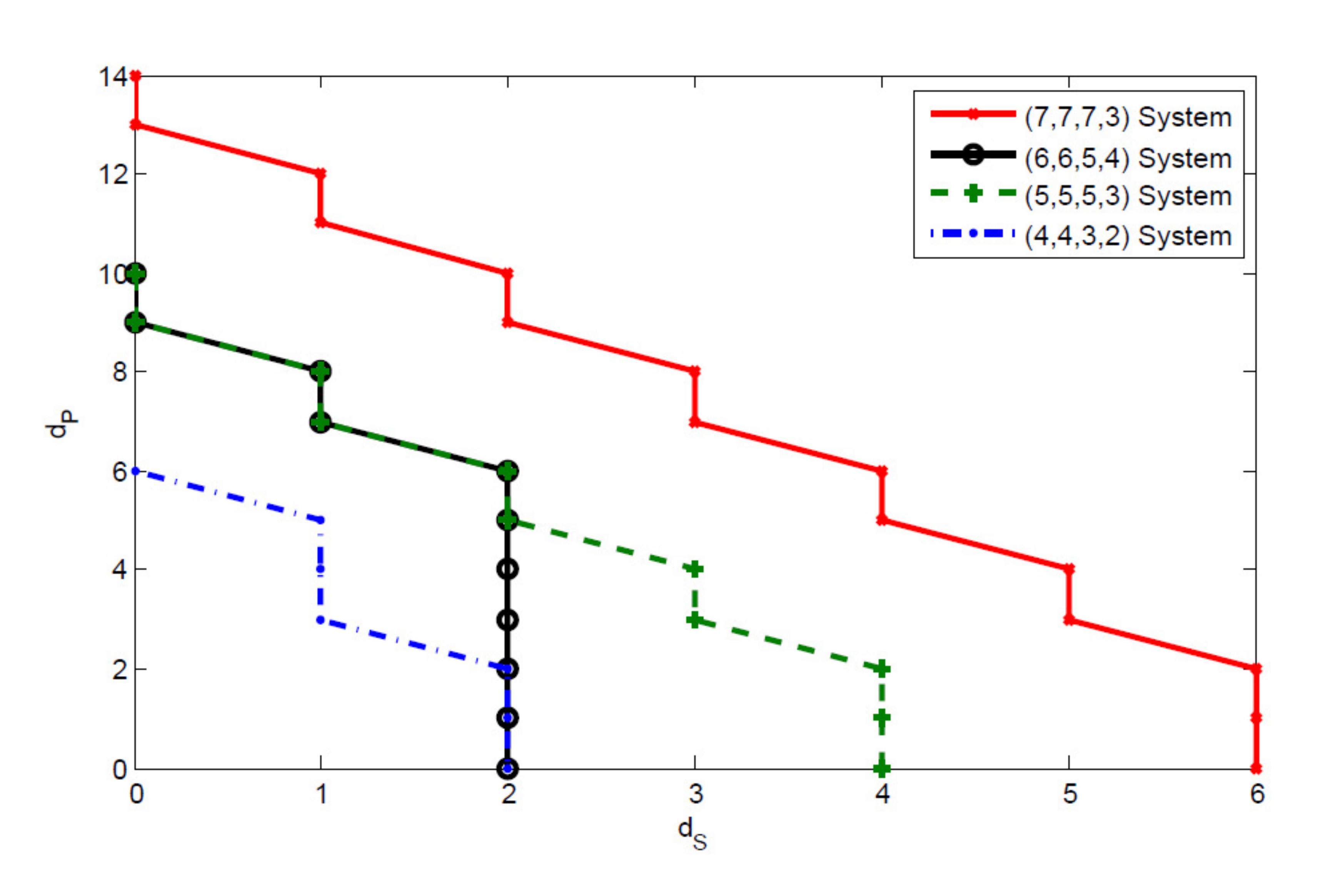}
  \caption{Achievable DoF regions for different types of CRNs.}
  \label{Fig:Dof}
\end{figure}

The interference cancellation scheme in ~\cite{Joint_IC} is first assumed. The DoF region is denoted as $\mathcal{D} = (d_{P_1},d_{P_2},d_{S_1},d_{S_2})\in\mathbb{R}^4$, where $d_{P_i}$ and $d_{S_j}$ are defined as the interference-free signalling dimensions for PUs $P_i$ and SUs $S_j$ respectively. The maximum sum DoF $d_m$ and the maximum sum DoF in the primary and secondary cells $d_P$ and $d_S$ can be found in ~\cite{Joint_IC}. In addition to the bounds presented in ~\cite{Joint_IC}, both $d_{S_1}$ and $d_{S_2}$ are bounded by the difference between the number of transmit and receive antennas in the sCell, that is $d_{S_1} \leq M_S-N_S$ and $d_{S_2} \leq M_S-N_S$.  This bound will be proved in the subsection below on interference cancellation of sCell. The number of available dimensions on each transmit and receive antenna is denoted with the quartet $({M_P,M_S,N_P,N_S})$ and provide the achievable region of the cognitive cellular network set-up. Fig. ~\ref{Fig:Dof} shows how the DoF of the system are affected by this quartet of values. It is also illustrated how the number of streams transmitted in each cell affects the possible number of interference-free streams that could be transmitted in the other cell. As illustrated, the bounds $d_{S_1}$ and $d_{S_2}$ on the horizontal axis clearly limits the DoF in the sCell.

\textit{Interference Cancellation at pCell}: We consider the primary transmission. Let $\textbf{x}^{P_1}$ and $\textbf{x}^{P_2}$ denote the [$d_{P_1} \times 1$] and [$d_{P_2} \times 1$] data streams intended to users $P_1$ and $P_2$, respectively. To construct $\textbf{x}_P$, $\textbf{x}^{P_1}$ and $\textbf{x}^{P_2}$ are processed through the [$M_P \times d_{P_1}$] and [$M_P \times d_{P_2}$] beam-forming/pre-coding matrices $\textbf{V}_{P_1}$ and $\textbf{V}_{P_2}$ respectively, and then added as follows:
\begin{equation} \label{Eq1}
\textbf{x}_P = \sum_{l=1}^{d_{P_1}}{\textbf{v}_{l}^{P_1} x_{l}^{P_1}}  + \sum_{k=1}^{d_{P_2}}{\textbf{v}_{k}^{P_2} x_{k}^{P_2}}, 
\end{equation}
where $\textbf{v}_{l}^{P_1}$ and $\textbf{v}_{k}^{P_2}$ are the $l^{th}$ and the $k^{th}$ columns of the matrices $\textbf{V}_{P_1}$ and $\textbf{V}_{P_2}$, respectively. The pre-coding matrices $\textbf{V}_{P_1}$ and $\textbf{V}_{P_2}$ are designed to cancel part of the intra-cell interference within the pCell corresponding to the dimension of the null space between $P$ and the PUs. We denote this dimension by $Z = (M_P-N_P)^{+}$. The first $Z$ columns of $\textbf{V}_{P_1}$ and $\textbf{V}_{P_2}$, i.e. $\textbf{v}_{1}^{P_1},...,\textbf{v}_{Z}^{P_1}$ and $\textbf{v}_{1}^{P_2},...,\textbf{v}_{Z}^{P_2}$, are chosen in the null space of $\textbf{H}_{P_2}$ and $\textbf{H}_{P_1}$, respectively. The rest of the ($d_{P_1} - Z$) and ($d_{P_2} - Z$) vectors of the matrices $\textbf{V}_{P_1}$ and $\textbf{V}_{P_2}$, respectively, are randomly selected with elements chosen according to an isotropic distribution.

The signal transmitted by  $S$ is designed to cancel the remaining part of the intra-cell interference within the pCell as follows:
\begin{eqnarray}
\nonumber
\textbf{x}_S &=& \sum_{g=1}^{d_{S_1}}{\textbf{v}_{g}^{S_1} \hat{x}_{g}^{S_1}} + \sum_{h=1}^{d_{S_2}}{\textbf{v}_{h}^{S_2} \hat{x}_{h}^{S_2}} +\\
\label{Eq2}
& & \sum_{l=Z+1}^{d_{P_1}}{\bar{\textbf{v}}_{l}^{P_1} x_{l}^{P_1}} + \sum_{k=Z+1}^{d_{P_2}}{\bar{\textbf{v}}_{k}^{P_2} x_{k}^{P_2}},
\end{eqnarray}
where $\hat{\textbf{x}}^{S_1}$ and $\hat{\textbf{x}}^{S_2}$ are the DPC-encoded data streams intended for users $S_1$ and $S_2$, respectively. $\textbf{v}_{h}^{S_1}$ and $\textbf{v}_{g}^{S_2}$ are the $h^{th}$ and $g^{th}$ columns of the [$M_S \times d_{S_1}$] and [$M_S \times d_{S_2}$] pre-coding matrices $\textbf{V}_{S_1}$ and $\textbf{V}_{S_2}$, respectively.

The signal received at $P_1$ will be given by:
\begin{eqnarray}
\nonumber
\textbf{y}_{P_1} &=& \sum_{l=1}^{Z}{\textbf{H}_{P_1}\textbf{v}_{l}^{P_1}x_{l}^{P_1}} + \sum_{l=Z+1}^{d_{P_1}}{\left( \textbf{H}_{P_1}-\textbf{H}'_{P_1}{\textbf{H}_{P_2}'}^{T}\right.}\\
\nonumber
& &\left.{\left(\textbf{H}'_{P_2}(\textbf{H}_{P_2}')^{T}\right)}^{-1} \textbf{H}_{P_2}\right)\textbf{v}_{l}^{P_1}x_{l}^{P_1} + \\
& &\textbf{H}'_{P_1}\left(\sum_{g=1}^{d_{S_1}}{\textbf{v}_{g}^{S_1}\hat{x}_{g}^{S_1}} +\sum_{h=1}^{d_{S_2}}{\textbf{v}_{h}^{S_2}\hat{x}_{h}^{S_2}}\right)+\textbf{z}_{P_1}.
\label{Eq3}
\end{eqnarray}
and the signal received at $P_2$ can be similarly defined. As shown in \eqref{Eq2}, the intra-cell interference in the pCell is totally cancelled using the pre-coding matrices.
 
To recover their data, $P_1$ and $P_2$ pass their received signals through the [$N_{P} \times d_{P_1}$] and [$N_{P} \times d_{P_2}$] post-processing matrices, $U_{P_1}$ and $U_{P_2}$. These post-processing matrices are formed as follows: For the user $P_1$ to decode $x_{l}^{P_1}$, it picks a zero-forcing vector, $\textbf{u}_{l}^{P_1}$, orthonormal to the space not containing $\textbf{v}_l^{P_1}$ for $1 \leqslant l \leqslant d_{P_1}$. The user $P_2$ picks the columns of $U_{P_2}$ similarly. 

\textit{Interference Cancellation at sCell side: } Here, we describe the secondary transmission scheme. The signal received at $S_1$, is given by: 
\begin{eqnarray}
\nonumber
\textbf{y}_{S_1} &=& \textbf{H}_{S_1}\left[ \sum_{g=1}^{d_{S_1}}{\textbf{v}_{g}^{S_1} \hat{x}_{g}^{S_1}} + \sum_{h=1}^{d_{S_2}}{\textbf{v}_{h}^{S_2} \hat{x}_{h}^{S_2}} \right.\\
\nonumber
&+& \left. \sum_{l=Z+1}^{d_{P_1}}{\bar{\textbf{v}}_{l}^{P_1} x_{l}^{P_1}} + \sum_{k=Z+1}^{d_{P_2}}{\bar{\textbf{v}}_{k}^{P_2} x_{k}^{P_2}}\right]\\
&+& \textbf{H}'_{S_1}\left[\sum_{l=1}^{d_{P_1}}{\textbf{v}_{l}^{P_1} x_{l}^{P_1}}  + \sum_{k=1}^{d_{P_2}}{\textbf{v}_{k}^{P_2} x_{k}^{P_2}}\right] +\textbf{z}_{S_1}.
\label{Eq4}
\end{eqnarray}
The signal received at $S_2$ can be similarly defined. The secondary's pre-coding vectors, $\textbf{V}_{S_1}$ and $\textbf{V}_{S_2}$ are designed to zero-force the intra-cell interference within the sCell. Without loss of generality, choose the first $d_{S_1}$ rows of $\textbf{H}_{S_1}$, $[(\textbf{h}^{1}_{S_1})^{T};...;(\textbf{h}^{d_{S_1}}_{S_1})^{T}]$ and the first $d_{S_2}$ rows of $\textbf{H}_{S_2}$, $[(\textbf{h}^{1}_{S_2})^{T};...;(\textbf{h}^{d_{S_2}}_{S_2})^{T}]$ to construct the following spaces:
\begin{eqnarray}
\mathcal{S}1 &=& [(\textbf{h}^{1}_{S_1})^{T};...;(\textbf{h}^{d_{S_1}}_{S_1})^{T}; \textbf{H}_{S_2}] \\
\mathcal{S}2 &=& [(\textbf{h}^{1}_{S_2})^{T};...;(\textbf{h}^{d_{S_2}}_{S_2})^{T}; \textbf{H}_{S_1}].
\label{Eq5}
\end{eqnarray}

The columns of $\textbf{V}_{S_1}$ and $\textbf{V}_{S_2}$ are selected as follows: for $1 \leqslant g \leqslant d_{S_1}$,  the $[M_S \times 1]$ vector $\textbf{v}_g^{S_1}$ will be picked in the null space of $\mathcal{S}1$ excluding the $g^{th}$ row. Similarly, for $1 \leqslant h \leqslant d_{S_2}$, the $[M_S \times 1]$ vector $\textbf{v}_h^{S_2}$ will be picked in the null space of $\mathcal{S}2$ excluding the $h^{th}$ row. Hence, the bound $d_{S_1} \leq M_S - N_S$ results \cite{Joint_IC}. 

\section{Achievable Sum Rates}
In this section, we analyse the achievable rate in the pCell and sCell.  

\textit{pCell rate} Let $\textbf{Q}^{P_i}$ denote the source power covariance matrix of the symbols in $\textbf{x}^{P_i}$ that is transmitted by $P$ and intended for user $P_i$.  We are interested in finding the source covariance matrix $\textbf{Q}^{P_i}$ that maximizes the achievable sum rate in the pCell, $R_{P}$, with an average power $Q_{av}$. This problem can be expressed as follows ~\cite{Nile_Journal}: 
\begin{eqnarray}\label{Eq7}
\nonumber
\smash{\displaystyle\max_{\textbf{Q}^{P_i}}}&&\frac{1}{2} \sum_{i=1}^{2}{{log_{2}\left(\textbf{I}_{d_{P_i}} + \frac{1}{\sigma_{P_i}^{2}}\textbf{U}_{P_i}^{T}[\textbf{H}\textbf{V}]_{P_i}^{*}\textbf{Q}^{P_i}[\textbf{H}\textbf{V}]_{P_i}^{*T}\textbf{U}_{P_i}\right)}}\\
\nonumber
s.t.&& \; \textbf{Q}^{P_i} \geq 0,\: for\; i\in\{1,2\}\\
& & \frac{1}{2} \sum_{i=1}^{2} tr\; \{\textbf{V}_{P_i}\textbf{Q}^{P_i}\textbf{V}_{P_i}^{T}\}\leq Q_{av}^{P},
\end{eqnarray}
where
\begin{eqnarray}\label{Eq8}
\nonumber
[\textbf{H}\textbf{V}]_{P_i}^{*} &=& \sum_{l=1}^{Z}{\textbf{H}_{P_i}\textbf{v}_{l}^{P_i}x_{l}^{P_i}} + \sum_{l=Z+1}^{d_{P_i}}{\left( \textbf{H}_{P_i}-\textbf{H}'_{P_i}{\textbf{H}_{P_j}'}^{T}\right.}\\
\nonumber
& &\left.{\left(\textbf{H}'_{P_j}(\textbf{H}_{P_j}')^{T}\right)}^{-1} \textbf{H}_{P_j}\right)\textbf{v}_{l}^{P_i}x_{l}^{P_i}\\
& &for\; i\in\{1,2\}, j\in\{1,2\}, j\neq i.
\end{eqnarray}
Equation \eqref{Eq7} is solved using water-filling algorithm that works as follows:
\begin{itemize} 
\item Find the the singular value decomposition (SVD) ${\bf{\Phi}}_{P_i} {\bf{\Gamma}}_{P_i} {\bf{\Psi}}_{P_i}^{T}$ of the matrix ${\textbf{U}}_{P_i}^{T} [{\textbf{H}}_{P_i} {\textbf{V}]^{*}}_{P_i}$ for $P_i$ with $\forall i\in\{1,2\}$. The term ${\bf{\Gamma}}_{P_i}$ is a $[d_{P_i} \times d_{P_i}]$ diagonal matrix that contains the singular values $\{{\gamma}_{l}\}^{d_{P_i}}_{l=1}$ representing the received signal-to-noise ratio (SNR) for each user.
\item Using the water-filling concept calculate the $[d_{P_i} \times d_{P_i}]$ matrices ${\hat{\textbf{Q}}}^{P_i}$ as follows:
\begin{equation} \label{Eq9} 
{\hat{\textbf{Q}}}^{P_i}_{l} = \left( \lambda - \frac{{\sigma}_{P_i}^{2}}{{{\gamma}_{l}^{P_i}}^{2}}\right), 
\: for\; i\in\{1,2\},l\in\{1,..,d_{P_1}\}, 
\end{equation}
where $\lambda$ is the common Lagrange multiplier chosen to satisfy the constraints in \eqref{Eq7}.
\item Find the optimal transmit power, ${\textbf{Q}}^{P_i}$, based on the result of the water-filling methodology and the received SNR using the right singular vector, ${\bf{\Psi}}_{P_i}$ as follows:
\begin{equation} \label{Eq10}
{\textbf{Q}}^{P_i} = {\bf{\Psi}}_{P_i} {\hat{\textbf{Q}}}^{P_i} {\bf{\Psi}}_{P_i}^{T}.
\end{equation}
\end{itemize}

\textit{sCell rate:} Similar to the analysis provided for the pCell, the source covariance matrix $\textbf{Q}^{S_j}$ that maximizes the achievable sum rate in the sCell, $R_{S}$,  can be found by solving: 
\begin{eqnarray} \label{Eq11}
\nonumber
\smash{\displaystyle\max_{\textbf{Q}^{S_j}}}&&\frac{1}{2} \sum_{j=1}^{2}{{log_{2}\left(\textbf{I}_{d_{S_j}} + \frac{1}{\sigma_{S_j}^{2}}\textbf{U}_{S_j}^{T}\textbf{H}_{S_j}\textbf{V}_{S_j}\textbf{Q}^{S_j}\textbf{V}_{S_j}^{T}\textbf{H}_{S_j}^{T}\textbf{U}_{S_j}\right)}}\\
\nonumber
s.t.&& \; \textbf{Q}^{S_j} \geq 0,\: for\; j\in\{1,2\}\\
& & \frac{1}{2} \sum_{j=1}^{2} tr\; \{\textbf{V}_{S_j}\textbf{Q}^{S_j}\textbf{V}_{S_j}^{T}\}\leq Q_{av}^{S}.
\end{eqnarray}
This is solved again using the methodology based on the water-filling algorithm that we described above.

Fig. 2 depicts the primary sum rate vs. the secondary sum rate of a (5,5,5,3) CRN for different values of average transmit power $Q_{av}$. The figure shows how the achievable sum rates in both cells change with the different number of streams transmitted in each cell with at least one stream transmitted in each cell. Obviously, there is a symmetry on the achievable rates with almost identical number of streams (dot dashed line), and a higher capacity for the pCell (solid line) or sCell (dashed line) respectively when the number of streams are getting higher to their own cells.  

\begin{figure}  \label{Rate_Result}
  \centering
  \includegraphics[width=0.45 \textwidth, height=0.25 \textheight]{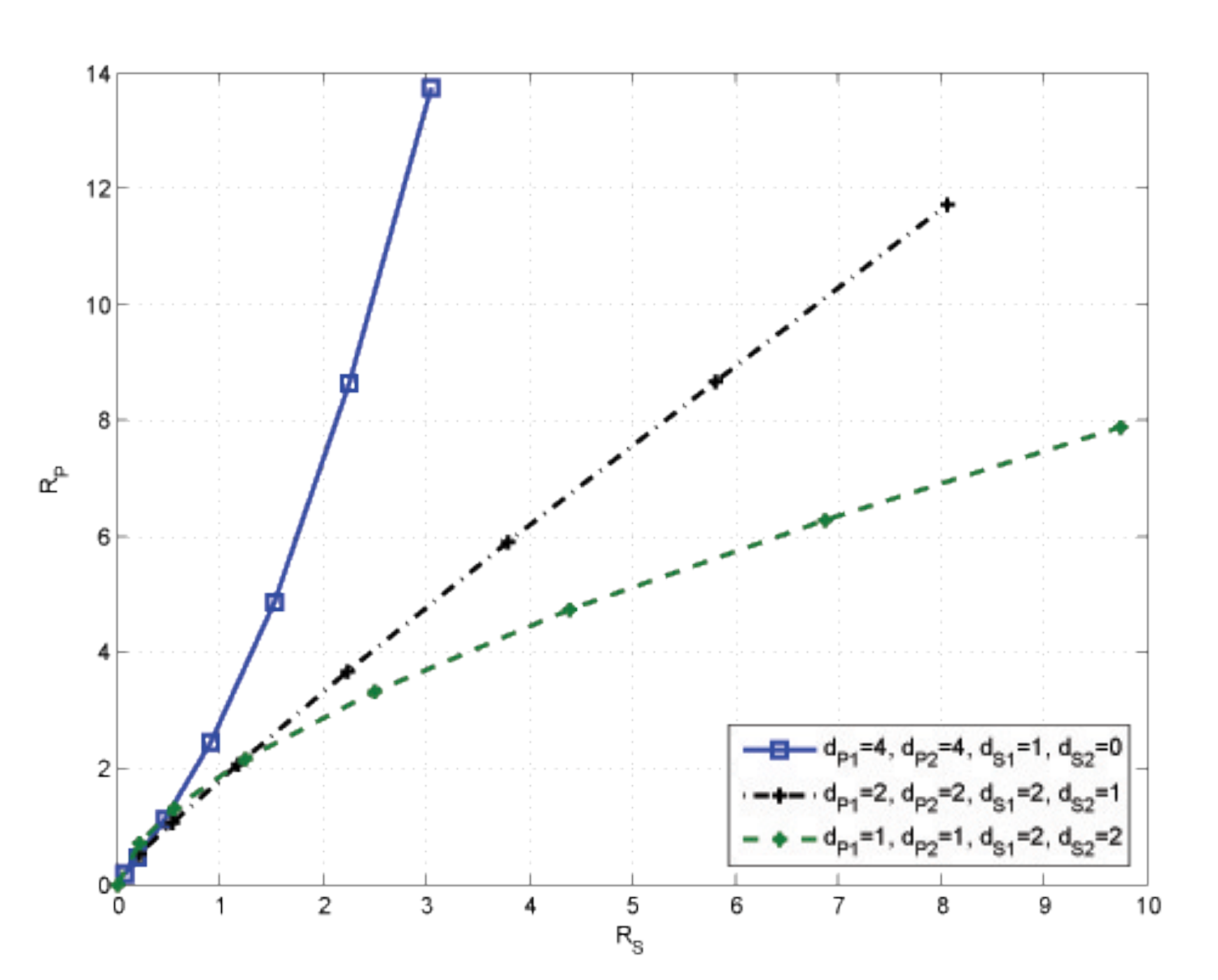}
  \caption{Primary sum Rate versus secondary sum Rate of a (5,5,5,3) CRN.}
\end{figure} 

\section{Conclusion}
We have investigated an interference alignment technique for the downlink of cognitive cellular network that allows complete interference cancellation and efficient utilization of the available DoF. We present details about the signalling and the achievable DoF regions and next derive the analysis for the rate maximization within each cell. The results of the achievable degrees of freedom and sum rate within each cell highlight the performance of the proposed scheme.

\vskip3pt
\ack{This work has been supported by QNRF project with code number: NPRP 09-1168-2-455}

\vskip5pt

\noindent M. Moussa, F. Foukalas and T. Khattab (\textit{Electrical Engineering, Qatar University, Doha, Qatar})
\vskip3pt

\noindent E-mail: foukalas@qu.edu.qa

\end{document}